\documentclass[namedreferences]{kluwer}
\usepackage{klups}
\usepackage[dvips]{graphicx}
\usepackage{makeidx}

\begin{document}
\begin{article}
\begin{opening}
\title{The UV Upturn: From M32 to Distant Clusters}
%\subtitle{}

\author{Thomas M. \surname{Brown}\email{tbrown@stsci.edu}}
\institute{Space Telescope Science Institute, 3700 San Martin Drive,
Baltimore, MD 21218, USA}                               

%\date:
%\dedication{}
%\translation{}

\runningtitle{The UV Upturn: From M32 to Distant Clusters}
\runningauthor{Brown}

%\begin{ao}
%\end{ao} 
%\begin{motto}
%\end{motto}

\begin{abstract} 
I review the observational constraints on the stars responsible for
the upturn in the UV spectra of ellipticals, ranging from galaxies in
the local Universe to distant clusters.  In nearby galaxies, this UV
upturn is produced by a minority population of extreme horizontal
branch (EHB) stars, with the large variations observed in the
UV-to-optical flux ratio driven by variations in the number of EHB
stars, and not the type of UV-bright stars.  Deep UV images of the
nearest elliptical galaxy, M32, show that it has a well-populated EHB,
even though it has the weakest UV upturn of any known elliptical
galaxy.  However, M32 suffers from a striking dearth of the hot
post-HB stars expected from canonical evolutionary theory.  As we
observe to larger lookback times in more distant galaxy clusters, the
UV upturn fades, as predicted by theories of stellar and galactic
evolution, but does so gradually.  Because the EHB stars do not appear
suddenly in the Universe, their presence is likely driven by a large
dispersion in the parameters that govern HB morphology.
\end{abstract}

%\keywords{}
%\abbreviations{\abbrev{}{}; \abbrev{}{}}
%\nomenclature{\nomen{}{}; \nomen{}{}}
%\classification{}{}
\end{opening}

\section{Introduction}

The source of UV emission from elliptical galaxies was one of the
great mysteries of extragalactic astrophysics for nearly 30 years.
The ``UV upturn'' manifests itself as a rising flux shortward of
2500~\AA.  Although the UV upturn is usually associated with the
spectra of elliptical galaxies, it was actually discovered in the
bulge of the nearby spiral M31 \cite{oao2}.  Prior to those
observations, spiral bulges and elliptical galaxies were thought to
contain only cool, passively-evolving populations of old stars.
Although elliptical galaxies have very similar optical spectra, their
UV-to-optical flux ratios, as measured by the $m_{1550}-V$ color
index, show strong variations, correlated with metallicity
\cite{mg2uvx}, such that galaxies with higher metallicity (optical
Mg$_2$ index) are bluer.  By 1990, there were many candidates for the
source of the UV emission, including young massive stars, binaries,
hot white dwarfs, extreme horizontal branch (EHB) stars,
post-asymptotic giant branch (post-AGB) stars, and non-thermal
activity \cite{GR90}.  Arguments based upon the fuel consumption
during different evolutionary phases made EHB stars a likely source,
one that implied a strong decline in the UV upturn at increasing
redshift ($z$) \cite{GR90}.

In the past decade, UV observations of elliptical galaxies in both the
local and distant Universe have proved conclusively that EHB stars are
the source of the UV upturn, and mapped its evolution over the range
$0 \le z \le 0.6$.  In these proceedings, I review this observational
evidence.  Section 2 reviews the spectroscopic observations of nearby
elliptical galaxies, which are well-matched by the integrated light of
EHB stars and their descendents. Section 3 covers the UV imaging of
the nearest elliptical galaxy, M32, which resolves the EHB population
responsible for the UV upturn but shows a surprising scarcity of the
UV-bright stars expected in the later stages of stellar evolution.
Section 4 discusses the evolution of the UV upturn with redshift.

\section{Spectroscopy of Nearby Elliptical Galaxies}

During the Astro-1 and Astro-2 missions, the Hopkins Ultraviolet
Telescope (HUT) observed six quiescent elliptical galaxies spanning a
wide range in $m_{1550} - V$ (\opencite{F91}; \opencite{hut1};
\opencite{hut2}).  The fast focal ratio, large apertures, and
wavelength coverage down to the Lyman limit made HUT an ideal
instrument for observing extended objects and for determining the
effective temperatures of hot UV sources.  The HUT spectra (figure 1)
were inconsistent with young massive stars, having a lack of strong
C\textsc{iv} $\lambda\lambda 1548,1551$ absorption and a declining
continuum from 1050~\AA\ to the Lyman limit (i.e., implying
temperatures less than 25,000~K for the UV sources).  Instead, the
spectra were well-matched by the integrated light expected from
populations of EHB stars and their descendents.

\begin{figure}
\centerline{\includegraphics[width=4.5in]{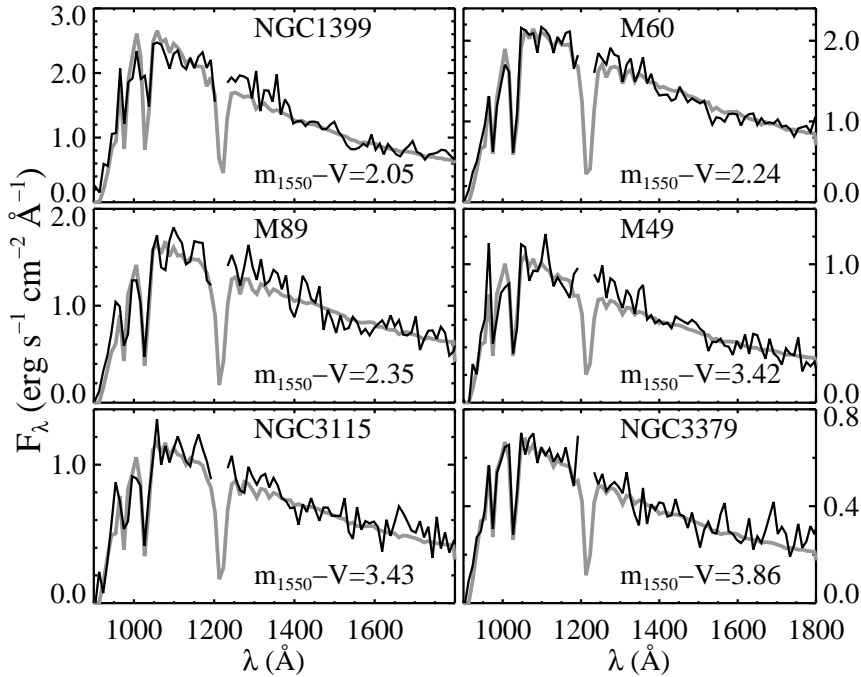}}
\caption{Spectra of six elliptical galaxies observed with the Hopkins
Ultraviolet Telescope ({\it black curves}).  Although the galaxies
span a large range in $m_{1550}-V$ (labeled), they all appear very similar,
and are well-matched by the integrated light ({\it grey curves}) 
of EHB stars and their progeny, spanning a narrow range of envelope mass.}
\end{figure}

\begin{figure}
\centerline{\includegraphics[width=4.5in]{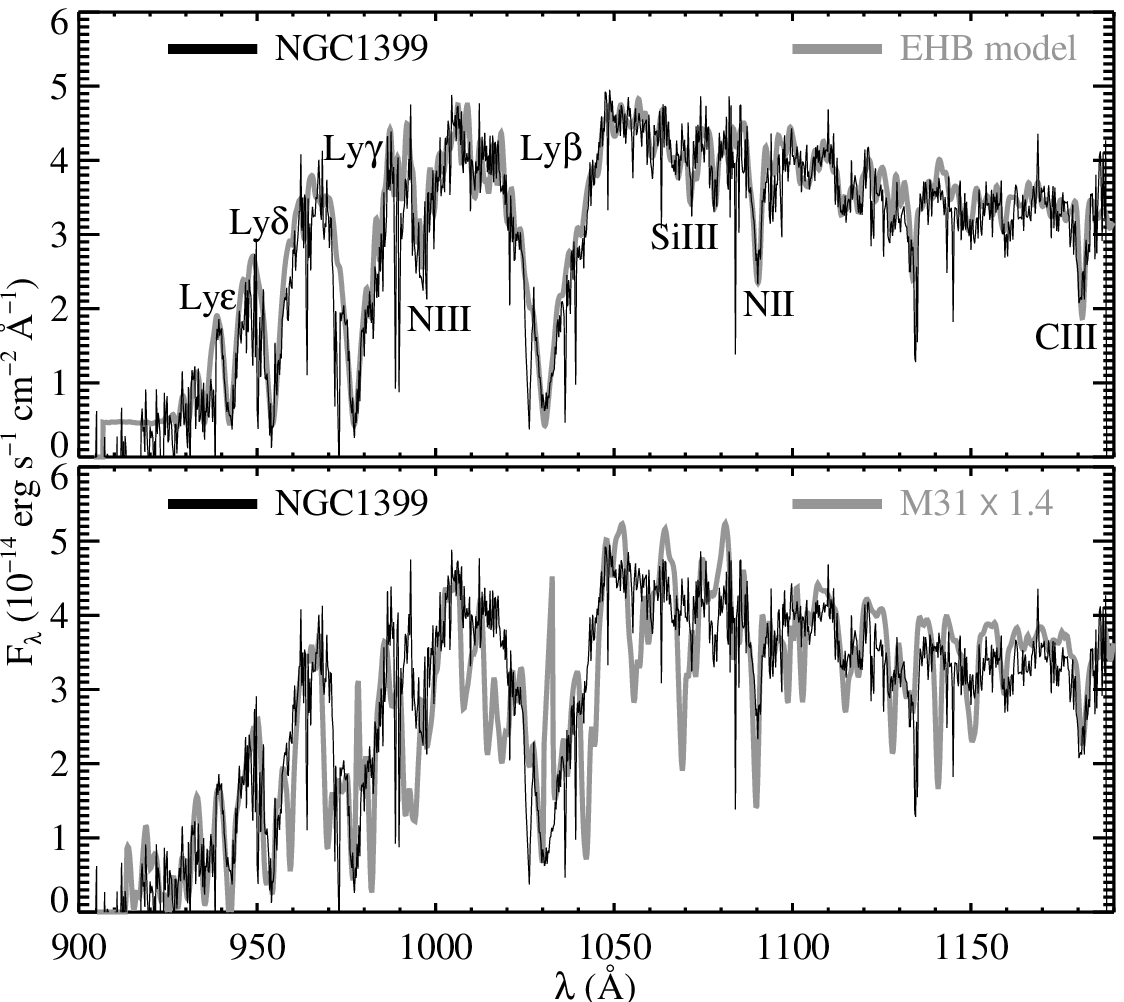}}
\caption{{\it Top panel:} The FUSE spectrum of NGC1399 ({\it black
curve}) is compared to the best-fit EHB model ({\it grey curve}) from
the earlier analysis of HUT data.  Photospheric features are labeled;
note the sharp Galactic interstellar features to the blue of several
photospheric features.  {\it Bottom panel:} The NGC1399 spectrum ({\it
black curve}) is compared to the renormalized M31 spectrum ({\it grey
curve}), convolved to the NGC1399 velocity dispersion.  Although
strong interstellar features complicate the analysis of the M31
spectrum, the C\textsc{iii} $\lambda$1175 feature appears somewhat
weaker in M31 than in NGC1399.}
\end{figure}

Although the galaxies in figure 1 span nearly 2~mag in $m_{1550}-V$,
their spectra are very similar, and well fit by models with a narrow
range in envelope mass on the EHB (0.02--0.09 $M_\odot$).  Neither
post-AGB stars (descendents of red HB stars) or post-early AGB stars
(descendents of blue HB stars) can contribute significantly to any of
these spectra, because their spectra are respectively hotter and
cooler than those observed.  This demonstrates that the strong
variations in the UV emission, relative to the optical, are the result
of variations in the fraction of EHB stars in the population, and not
a variation in the type of stars producing the UV flux.  Moreover, the
HB distribution in each galaxy must be strongly bimodal, with a
significant but minority ($\lesssim 10$\%) population of EHB stars, very few
blue HB stars, and a majority population of red HB stars.

The successor to HUT, the Far Ultraviolet Spectroscopic Explorer
(FUSE), has been operating since 1999.  Although it observed the giant
elliptical galaxy NGC1399 (in Fornax) early in the mission
\cite{n1399}, reaction wheel failures have rendered the Virgo cluster,
where nearly all nearby elliptical galaxies reside, virtually
unobservable.  Thus, observations of M60 were never completed, and
observations of additional ellipticals are unlikely.  Figure 2 shows
the NGC1399 spectrum and the same best-fit EHB model that matched the
HUT spectrum, but at higher resolution.  Although the resolution for
FUSE (0.025~\AA) is much higher than that of HUT (3~\AA), the
velocity dispersion in NGC1399 effectively limits the resolution to
$\sim$1~\AA.  Nevertheless, the increase in resolution and
signal-to-noise allows a determination of the photospheric abundances
for the EHB stars driving the UV upturn.  The C abundance is 2\%
solar, the Si abundance is 13\% solar, and the N abundance is 45\%
solar.

The UV upturn is positively correlated with optical metallicity
indicators, contradicting the general tendency for the HB morphology
to become redder at increasing metallicity.  This has led to
considerable debate about the metallicity of the hot stars responsible
for the UV emission.  In particular, \inlinecite{PL97} have argued that the
UV upturn should be anticorrelated with the metallicity of the hot
stars.  In their view, the correlation between UV upturn and optical
metallicity indicators is due to the more metal-rich galaxies being
older and more massive.  Others (e.g., \opencite{GR90};
\opencite{BCF94}) have argued that the tendency for redder HB
morphology at increasing metallicity is reversed at high
metallicities, because of an associated increase in helium abundance
and perhaps enhanced mass loss; in this case, the UV upturn should
positively correlate with both optical and UV metallicity indicators.
The metallicity of the EHB stars in NGC1399, the galaxy with the
strongest known UV upturn, is clearly neither metal-rich nor
metal-poor.  

The surface abundance pattern of the EHB stars, derived from the UV
spectra of elliptical galaxies, is probably affected by diffusion in
the stellar atmospheres, as often found for sdB stars in the Galactic
field.  Nevertheless, it is still interesting to compare the
metallicity in UV-strong galaxies with UV-weak galaxies, to look for
general trends.  Unfortunately, FUSE did not observe the UV-weak giant
elliptical M49 before the onset of its hardware problems, and the
spectrum of the nearby galaxy M32 is of very low signal-to-noise.
Given the inability to point at Virgo, we replaced M49 in our program
with the bulge of M31, which also has a fairly weak UV upturn.  Its
spectrum is shown in figure 2; because the bulge of M31 does not have
the high velocity dispersion of NGC1399, the M31 spectrum has been
convolved and renormalized appropriately for comparison.  The spectrum
of M31 is much more difficult to interpret than that of NGC1399; the
redshift of NGC1399 separates the weak Galactic interstellar lines
from the EHB photospheric lines, while the slight blueshift of M31 is
not enough to separate these lines.  Unlike NGC1399, which suffers
from no foreground extinction, the foreground extinction toward the
M31 nucleus is much stronger: $E(B-V)=0.08$~mag \cite{ext}.
Nevertheless, a few photospheric features have no interstellar
counterpart, such as C\textsc{iii} $\lambda$1175.  This feature shows
that the C abundance in M31 is also very weak -- perhaps weaker than
that in NGC1399.  In general, the continuum shape is very similar in
the two spectra.  However, after dereddening, the M31 spectrum would
appear somewhat hotter than the NGC1399 spectrum, suggesting a
contribution from either hotter EHB stars and/or a higher contribution
from post-AGB stars.

\section{UV Images of M32}
Because M32 has the weakest known UV upturn \cite{mg2uvx}, its IUE
spectrum could have been explained entirely by relatively short-lived
post-AGB stars instead of EHB stars.  Early UV images of M32 and the
M31 bulge, with the Faint Object Camera (FOC) on the Hubble Space
Telescope (HST), resolved some of this UV emission into bright stars
\cite{m31m32}, but later near-UV images with the Space Telescope
Imaging Spectrograph (STIS) resolved the EHB stars in M32,
demonstrating that they are responsible for nearly all of its UV
emission \cite{m32}.  Although no color information was available in
those data, the near-UV luminosity function showed a sharp peak at the
level of the EHB.  Recently, we obtained far-UV observations of the
same field with the HST/STIS, in order to construct a deep UV
color-magnitude diagram (CMD) of M32 (figure 3).  The CMD shows a
well-populated EHB, confirming that the UV emission comes from a
minority population of EHB stars ($\sim 2$\% of the total HB).  However,
the images are missing many of the post-HB stars expected from canonical
stellar evolution theory.  Most of the HB stars in M32 lie on the red
HB, and those should produce hundreds of UV-bright post-AGB stars, yet
the HST/STIS images have only a handful.  Furthermore, there should be
approximately 1--2 post-EHB stars (also known as AGB-Manqu$\acute{\rm
e}$ stars) for every 10 EHB stars, yet these post-EHB stars are also
under-represented.  This is shown by the simulation in figure 3, which
assumes a bimodal HB morphology that best reproduces the distribution
of stars in the observed CMD.  Although the number of EHB stars can be
matched, no mass distribution reproduces the number of
AGB-Manqu$\acute{\rm e}$ and post-AGB stars.  Because the missing
stars should be the brightest ones in the image, they cannot be
missing due to instrumental effects or incompleteness.  Instead, these
post-HB stars likely evolve on much more rapid timescales than
predicted by standard evolution theory.

\begin{figure}
\centerline{\includegraphics[width=4.5in]{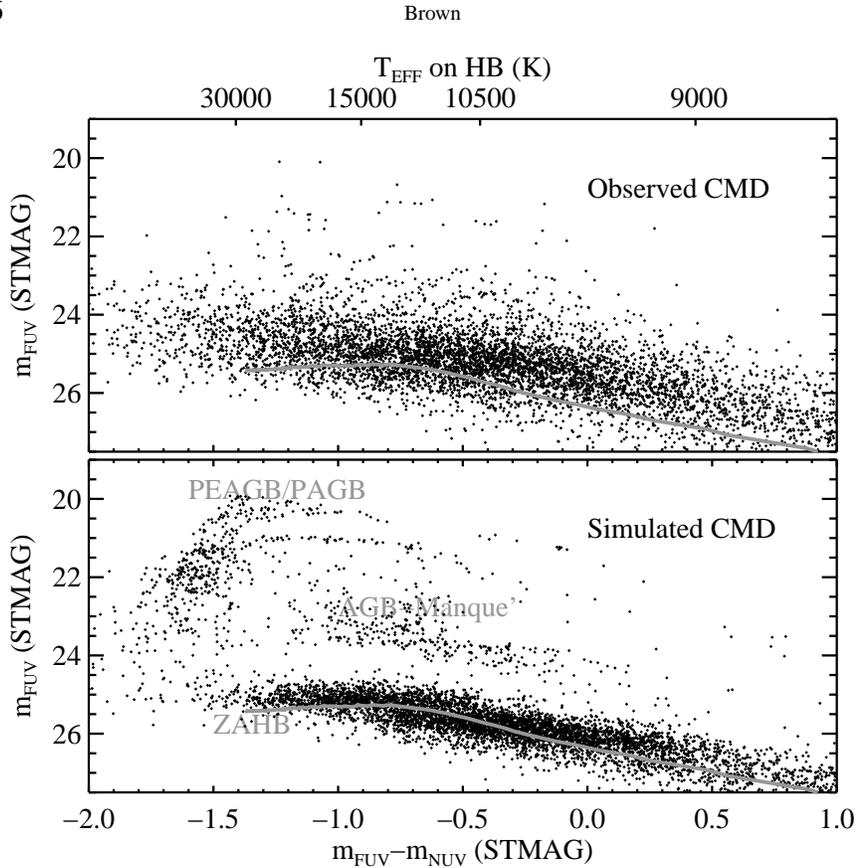}}
\caption{{\it Top panel:} The CMD constructed from the near-UV and
far-UV images of M32, as observed by the HST/STIS ({\it black
points}), compared to the predicted location of the zero-age HB ({\it
grey curve}).  Although the EHB is well-populated, the UV-bright stars
above the EHB are under-represented.  {\it Bottom panel:} The
simulation that best reproduces the observed CMD, which assumes a
bimodal HB morphology having a minority EHB population.  Note the
large numbers of UV-bright stars, and the clear gap between the EHB
and AGB-Manqu$\acute{\rm e}$ stars, which are not seen in the observed
CMD.}
\end{figure}

\section{The Evolution of the UV Upturn}
Because EHB stars are the source of the UV upturn, the UV upturn is
expected to fade dramatically with increasing redshift as one looks to
younger elliptical galaxies (\opencite{GR90}; \opencite{T96}).  We
have been undertaking a series of observations with HST to map the
evolution of the UV upturn as a function of redshift, by observing
galaxy clusters at $0.3 < z < 0.6$ (figure 4).  Although early
measurements were consistent with a relatively flat evolution out to
$z \approx 0.4$ and a rapid fading at higher redshifts
(\opencite{z375}; \opencite{z55}), recent observations \cite{z33} at
$z=0.33$ show a UV upturn as weak as that at $z=0.55$.  Because the
earliest observations, at $z=0.375$, were the only ones that did not
use a solar-blind camera, they might be systematically in error;
setting aside the $z=0.375$ measurements, the remaining observations
show a UV upturn that is weaker than that in the present epoch, but a
relatively flat evolution with increasing lookback time. Taking the
models at face value implies large variations in the formation epoch
of giant elliptical galaxies in clusters, which is implausible
\cite{z33}. Although the onset of the UV upturn occurs at $\sim$6 Gyr
in these models, the formation of EHB stars is tied to a wide range of
poorly constrained parameters (mass loss, metallicity, binary
fraction, etc.), with the elliptical galaxies presenting the aggregate
behavior at any one epoch.  The ``floor'' in the UV emission seen at
increasing redshift might indicate a wide dispersion in the parameters
that govern EHB formation, or it might be that another source of UV
emission is becoming dominant at increasing redshift as the EHB stars
disappear (e.g., residual star formation).  With our current understanding
of EHB formation, the UV upturn remains a poorly calibrated age indicator.
However, if galaxy ages are determined by independent methods, these
surveys of the UV upturn could instead be used to constrain theories
of EHB formation.

\begin{figure}
\centerline{\includegraphics[width=3.9in]{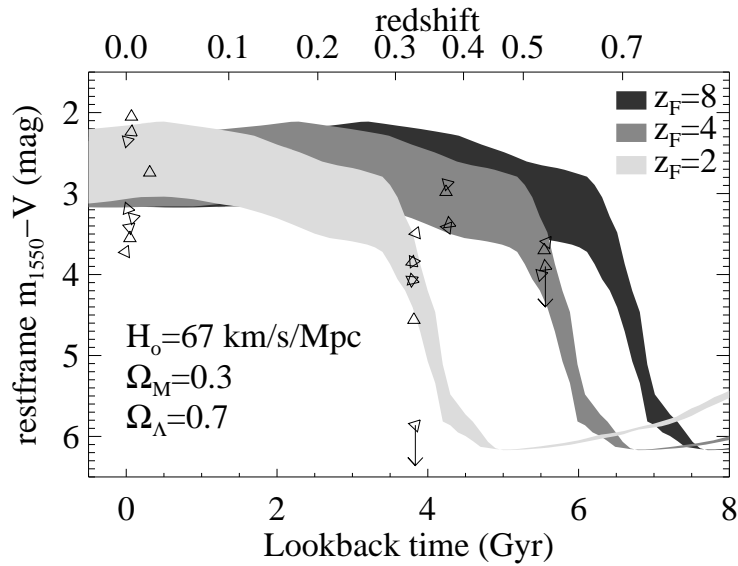}}
\caption{
The evolution of the UV upturn as a function of redshift
({\it triangles}), with the expected evolution of this emission in
giant elliptical galaxies \protect\cite{T96},
assuming three different formation redshifts ($z_f$; {\it labeled}).}
\end{figure}

\section{Summary}
The past decade of observations in nearby galaxies has shown that EHB
stars are the dominant source of the UV upturn, even in those galaxies
with very weak UV emission, such as M32.  We can now resolve EHB stars
in galaxies beyond the Milky Way, out to the distance of M31 and its
satellites.  UV images of M32 show a well-populated EHB, but a
surprising dearth of AGB-Manqu$\acute{\rm e}$ and post-AGB stars.  As
we look to elliptical galaxies at higher redshift, the UV upturn
fades, but not as rapidly as might be expected, suggesting either a
large dispersion in the parameters that govern the formation of EHB
stars, or another source of UV emission that becomes dominant at
earlier ages.  Although the UV upturn may be the most sensitive
indicator of age in an evolved population, it is a diagnostic that is
poorly constrained by our current understanding of EHB stars.

\begin{acknowledgements}
The work presented herein was done in collaboration with A. Davidsen
(JHU), H. Ferguson, R. Jedrzejewski, E. Smith (STScI), C. Bowers,
B. Dorman, R. Kimble, R. Ohl, A. Sweigart (NASA/GSFC), R.M. Rich (UCLA),
A. Renzini (ESO), J.-M. Deharveng (Laboratoire d'Astronomie Spatiale),
R. O'Connell (U. of Virginia), and S.A. Stanford (LLNL).  These observations
were supported by NASA contract NAS 5-27000 to the Johns Hopkins University,
NASA grant NAS 5-9696 to the Catholic University of America, NASA grant
NAS 5-6499D to the Goddard Space Flight Center, and NASA grants
NAG5-12278 and NAS 5-26555 to the Space Telescope Science Institute.
\end{acknowledgements}

\end{article}

\end{document}